# CsCuCl₃ perovskite-like compound under extreme conditions

J. S. Rodríguez-Hernández, Mayra. A. P. Gómez, O. P. Furtado, D. L. M. Vasconcelos,

A. P. Ayala, C. W. A Paschoal

Departamento de Física, Universidade Federal do Ceará

Leonardo O. Kutelak, Gustavo A. Lombardi, Ricardo D. dos Reis

Brazilian Synchrotron Light Laboratory (LNLS), Brazilian Center for Research in

Energy and Materials (CNPEM), Campinas, SP, Brazil

## Abstract

Halide perovskite has attracted intense research interest owing to its multifaceted and versatile applications in optoelectronics. This intrigue is further fueled by their propensity to undergo intricate structural modifications under extreme conditions, thereby instigating property changes. Within this context, our study delves deep into the intricate interplay of structural and vibrational attributes within the inorganic-metal halide perovskite-like CsCuCl₃. Our approach employs Raman spectroscopy and Synchrotron Powder X-Ray Diffraction (SPXRD) techniques harnessed under the dual conditions of low temperatures and high pressures. We have observed a distinct spin-phonon coupling mechanism by employing Raman spectroscopy at low temperatures; this coupling has been manifested as a renormalization phonon phenomenon that occurs notably at $T^* = 15$ K. The correlation between spin and phonon dynamics becomes pronounced through a notable hardening of phonon temperature dependence, a behavior intricately linked to the material antiferromagnetic transition at $T_N = 10.7$ K. The SPXRD under high pressure showed a first-order structural phase transition (SPT) at the critical pressure $P_c = 3.69$ GPa, leading to the transformation from the hexagonal $P6_522$ to a base-centered monoclinic cell. Notably, the coexistence of both phases is discernible within the pressure range from 2.79 to 3.57 GPa, indicating that the SPT involves the reorganization of the internal [Cu₂Cl₉]⁵⁻ dimer unit, with the Cl-Cu-Cl bending contributing more than stretching modes. Furthermore, we demonstrate that the SPT is reversible, but residual strain pressure influences the modification of the critical pressure $P_c$ value upon pressure decrease.

# I. INTRODUCTION

The $CsCuX_3$ family comprises a class of inorganic-metal halide perovskite-like compounds with a general chemical formula of $ABy_Xz$, where A = Rb, or Cs; B = Mn, Fe, Co, Ni, or Cu; and X = Cl, Br, or I. These materials have attracted significant attention due to their wide-ranging potential applications in optoelectronics, catalysis, and energy storage [1–4]. Notably, $CsCuX_3$ compounds exhibit optical and electronic properties like their Pb-based counterparts but offer advantages such as enhanced stability and reduced toxicity compared to lead halides. As a result, there has been growing interest in exploring partial or complete substitution of $Cu^+$ in Pb-based compounds [5–7].

Typically, halide perovskites with the general formula $ABX_3$ and a transition metal ion B adopt derivatives of the hexagonal perovskite-like structure, giving rise to compounds falling into the Ruddlesden-Popper and Dion Jacobsen series of materials [8,9]. From a symmetry perspective, the intrinsic *3-* or *6*-fold rotational symmetries lead to frustration in the ordering of magnetic moments or orbital occupancies of the framework ions into a single lowest-energy state at low temperatures, rendering them intriguing, promising candidates for achieving quantum spin-liquid state [10].

Among the $CsCuX_3$ family, $CsCuCl_3$ is a perovskite-like material with an antiferromagnetic order below $T_N = 10.7$ K ($Cu^{2+}$ - $S = \frac{1}{2}$) [11–13]. The antiferromagnetic (AFM) state arises from the exchange interaction within the intrachain and the antisymmetric exchange (Dzyaloshinskii-Moriya: DM) interaction, allowed by the the 120° twist in the AFM phases along the [001] direction [14]. At room temperature, the $CsCuCl_3$ adopts a polytype distorted hexagonal perovskite structure belonging to chiral space groups $P6_522$ or $P6_122$, which undergoes a structural phase transition at high temperatures (423 K) due to the Jahn-Teller effect, leading to the space group $P6_3/mmc$ [15]. The hexagonal phase $P6_522$ (commonly referred to as left-handed spin rotation direction) consists of two face-shared distorted $[CuCl_6]^{4-}$ octahedra forming a $[Cu_2Cl_9]^{5-}$ dimer unit, which displays a 1D chain of dimers along the *c*-axis, with the $Cs^+$ ion occupying the void space between the chains (See Figure S1 in the Supplementary Material).

The investigation of frustrated quantum many-body systems, such as $CsCuCl_3$, under high pressure provides an excellent opportunity to explore the effects of competing interactions at low-energy states. External pressure manipulation in frustrated quantum materials facilitates the active modulation of quantum correlations across classical and quantum-mechanical regimes, opening doors to studying exotic phenomena emerging in the crossover between these two regimes [17]. The magnetic diagram of CsCuCl$_3$ at low temperatures, as a function of the longitudinal magnetic field ($\boldsymbol{H} \parallel c$), exhibits a quantum-phase transition ($\boldsymbol{H}$ =12.5 T) from an umbrella phase to a *2-1* coplanar phase as the magnetic field increases. Similarly, applying pressure increases the incommensurate (IC) wavenumber with the magnetic field and pressure, enhancing neighboring spins and modifying the magnetic diagram through the enhanced Dzyaloshinskii-Moriya interaction [18,19]. Consequently, pressure-induced new quantum phases are observed, distinguished by the ICN notation (with N ranging from 1 to 5 for each phase), advancing the scientific understanding of controlling quantum mechanical correlations in weakly-coupled spin chain materials through external pressure in CsCuCl$_3$ [20].

Furthermore, the coupling between spin, charge, lattice, and orbital degrees of freedom is fundamental in condensed matter physics, giving rise to emergent phenomena and applications, such as multiferroics and spintronics [21]. Recent interest has grown in exploring spin-phonon coupling (SPC) in materials, which simultaneously control magnetic and phononic properties [22,23]. The stability of magnetic states can be influenced by modifying epitaxial strains or displacing magnetic ions in the sample through external conditions, such as high magnetic fields, high pressure, or low temperatures. The Hamiltonian governing spin-related phenomena in solids can be expressed as a sum of isotropic exchange (IE), Dzyaloshinskii-Moriya (DM), anisotropic exchange (AE), and single-ion anisotropy (SIA) interactions [24,25], given by:

$$H_{spin} = \sum_{ij}^{nn} [J_{IE}(\boldsymbol{S}_i \cdot \boldsymbol{S}_j) + \boldsymbol{D}_{ij} \cdot (\boldsymbol{S}_i \times \boldsymbol{S}_j) + \boldsymbol{S}_i \cdot \boldsymbol{\Gamma}_{ij} \cdot \boldsymbol{S}_j] + A \sum_i (\mathbf{n}_i \cdot \boldsymbol{S}_i)^2 \qquad (1)$$

Here, $J_{IE}$, $\boldsymbol{D}_{ij}$, $\boldsymbol{\Gamma}_{ij}$, and $A$ are the coefficients IE, DM, AE, and SIA interactions, respectively, while $\mathbf{n}_i$ is the vector direction axis, and $\boldsymbol{S}_i$, $\boldsymbol{S}_j$ denote neighboring spins at sites $i$ and $j$. These spin interactions are highly dependent on Coulomb interactions, but the terms $J_{IE}$, $\boldsymbol{D}_{ij}$, $\boldsymbol{\Gamma}_{ij}$, and $A$ are dynamically modified by lattice vibrations, leading to spin-phonon coupling (SPC) and the observed renormalization of phonon frequencies in certain cases [26].

Despite extensive investigations into the magnetic properties of $CsCuCl_3$ at low temperatures and high pressures, no systematic study of structural phase transitions or spin-phonon coupling in this crystal has been conducted. Therefore, this paper aims to bridge this gap by reporting on the presence of spin-phonon coupling in $CsCuCl_3$ during the low-temperature antiferromagnetic phase transition and a first-order high-pressure induced structural phase transition.

## II. EXPERIMENTAL AND COMPUTATIONAL METHODS

Dark red needles of $CsCuCl_3$ single crystals were grown through the slow evaporation method from a solution containing stoichiometric amounts of $CsCl$ and $CuCl_2$ in acidic HCl (47 wt% in $H_2O$). The solution was heated at 120 °C for 1 hour and then kept at room temperature into a becker sealed with paraffin film. The crystals were collected and cleaned with toluene.

The crystal structure was confirmed using Single Crystal X-Ray Diffraction (SCXRD) measurements performed on a Bruker D8 Venture X-ray diffractometer equipped with a Photon II Kappa detector and Mo K radiation ($\lambda = 0.71073$ Å) microfocus source. The crystal was chosen and mounted on a MiTeGen MicroMount using immersion oil. The obtained data were processed using APEX4 software [27] and Bruker SAINT software package for data reduction and global cell refinement, [28] respectively. The structure was solved by intrinsic phasing using SHELXT [29] and refined by least squares on SHELXL [30] included in Olex2 [31]. The crystallographic illustrations were prepared in MERCURY [32] and VESTA software [33]. The determined crystal structure belongs to the $P6_522$ space group with $a = 7.2168$ (1) Å and $c = 18.1853$ (5) Å as cell parameters, consistent with Chen *et al.* [16]. Some verified crystals were crushed to prepare a powder sample with 10 µm thickness, which was used for the measurements at extreme conditions.

Low-temperature Raman spectra of the powder sample were collected using a T64000 Jobin–Yvon spectrometer equipped with an Olympus microscope and an $LN_2$-cooled CCD to detect the scattered light. The spectra were excited with an Argon ion laser ($\lambda = 568$ nm), and the temperature-dependent spectra were obtained using a He-compressed closed-cycle cryostat with precise temperature control (± 0.1 K). A membrane diamond anvil cell (MDAC) chamber was utilized for pressure-dependent spectra, and Nujol served as the pressure-transmitting medium. The pressure was determined using the ruby pressure gauge [34]. The powder of $CsCuCl_3$ and ruby were together in the gaskets without contact. Each Raman spectrum was deconvoluted into the sum of Lorentzian functions using Fityk software [35].

The high-pressure Synchrotron powder XRD data were obtained at the Extreme Methods of Analysis (EMA) beamline of the Brazilian Synchrotron Light Laboratory. These measurements were conducted at ambient temperature, utilizing a monochromatic, 25.514 keV, X-ray beam. High pressure was generated using a diamond anvil cell (DAC) fitted with 600 μm diamond anvils. We used a stainless-steel gasket with a 200-micron hole into which the sample was loaded, along with a ruby ball.Neon was used as the pressure medium. Pressure was determined in-situ by measuring the wavelength of the ruby fluorescence second peak and was controlled via a gas-membrane mechanism integrated with the DAC. We employed an X-ray spot size of 15 μm × 15 μm at the sample position for the SRXD measurements. The two-dimensional diffraction images were captured in a transmission geometry using a MARCCD165 detector which has a pixel size of 73.2 μm × 73.2 μm. Subsequently, these images were processed and integrated using Dioptas 4.0 software [36] . To ensure accuracy, we calibrated the detector distance and other geometrical parameters using the NIST (National Institute of Standards and Technology, USA) standard reference material 660c ($LaB6$). The extracted XRD powder data was then refined using Rietveld and/or Lebail methods with EXPO 2014 [37].

Theoretical phonon calculations were performed using density functional theory (DFT) implemented into the Quantum-ESPRESSO package. The calculations included structural optimization using SG15 Optimized Norm-Conserving Vanderbilt pseudopotentials [40], followed by $\Gamma$-point phonon calculations. The phonon frequencies were calculated using the density functional perturbation theory (DFPT), with the exchange-correlation term determined within the generalized gradient approximation (GGA) parameterized by Perdev-Burke-Ernzerhof (PBE) [41]. Additionally, the local-density approximation (LDA) was employed for the exchange-correlation term for Raman intensity calculations.

# III. RESULTS AND DISCUSSIONS

## A. Typical Raman spectrum and phonon assignments

The group-theory analysis based on the site occupation of $CsCuCl_3$ yields 35 Raman-active phonon modes at room temperature, which are distributed among the irreducible representations of the point group *622* as the direct sum $\Gamma_{RAMAN} = 6A_1 \oplus 15E_2 \oplus 14E_1$ (see Table SI in the Supplementary Material). Figure 1(a) shows the Raman spectrum in the 100-320 cm$^{-1}$ range for the $CsCuCl_3$ powder sample at 9.2 K. Since $CsCuCl_3$ does not undergo any structural phase transition at low temperatures, this is the typical spectrum of $CsCuCl_3$ ($P6_522$ symmetry) (see Figure S2 in the Supplementary Material that shows the Raman spectrum at room temperature). The phonon assignments were based on the stable lattice dynamics calculated around the $\Gamma$-point using Differential Functional Perturbation Theory (DFPT). The DFPT calculations relax the crystallographic cell size and shape by minimizing all quantum forces in the static lattice, which approximates the crystal structure to $T = 0$ K.

Table I summarizes the experimental Raman (300 K and 9.2 K) and respective DFPT-calculated phonons, which were compared with previous Raman or IR measurements of similar materials such as $ABCl_3$ (A=Cs, Rb; B=Mn, Co) [42–44], $CsBBr_3$ (B=Co, Mg, Cd) [45], and $Cs_2XCl_4$ (X=Cu, Co) [46,47], $[(CH_3)_4N]_2MnX_4$ (X=Cl, Br) [48]. These phonons were separated into Cl-Cu-Cl bending and Cl-Cu-Cl stretching. Such calculated phonons are shown in Figure 1(b), which are described around the $[Cu_2Cl_9]^{5-}$ dimer unit. A full list of the calculated phonons is given in Table SII in Supplementary Material.

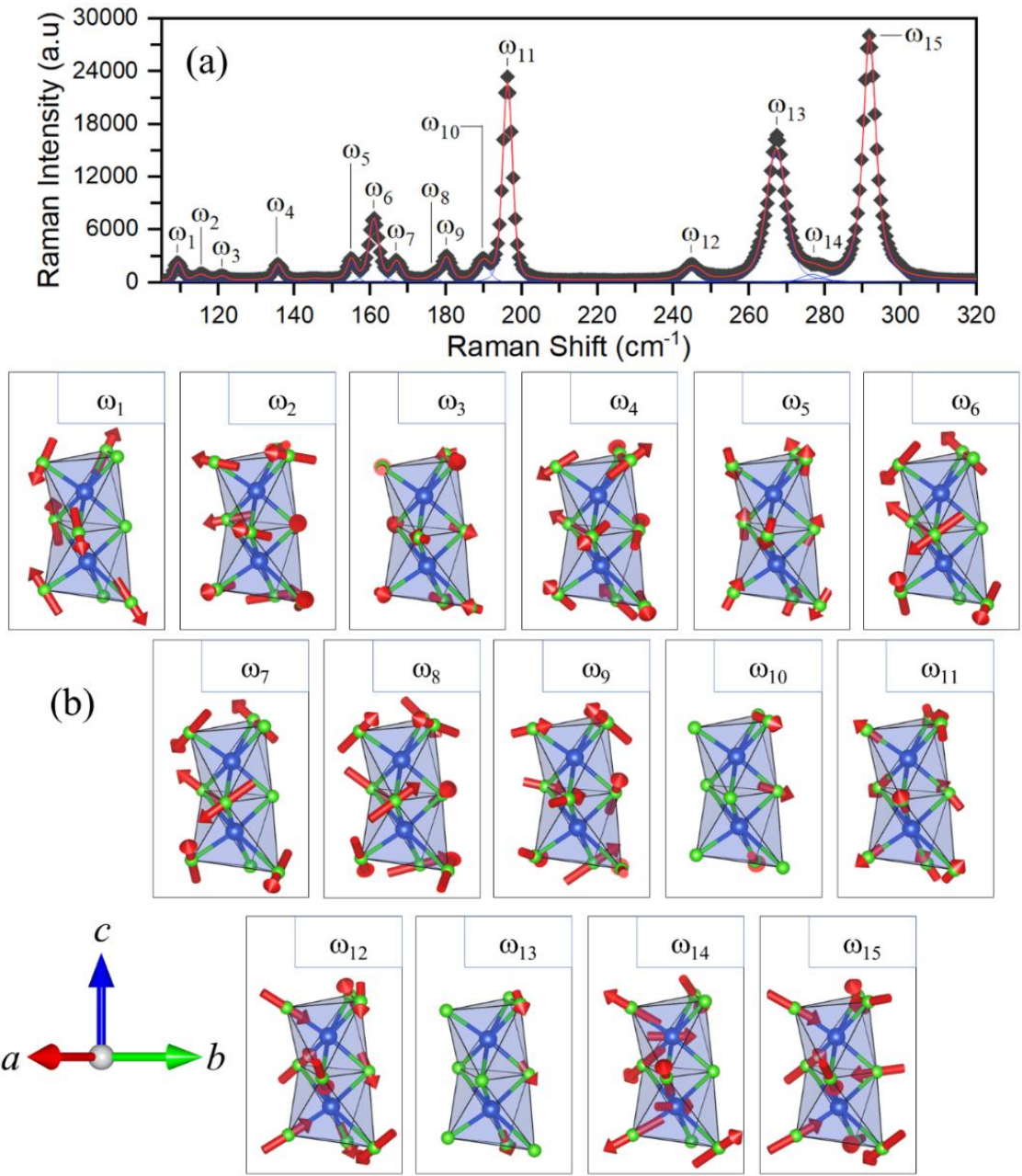

**Figure 1** – (a) Raman spectrum of CsCuCl$_3$ at 9.2 K. (The blue and red curves are the Lorentzian oscillator phonons bands and the total modes convolution, respectively). (b) Calculated Raman modes in CsCuCl$_3$. Note that the dimer unit [Cu$_2$Cl$_9$]$^{5-}$ is described along the *a*, *b* and *c* directions.

**Table I**. Experimental Raman modes (Exp.) at room-pressure (300K), low-temperature (9.2 K) and Density Functional Perturbation Theory (DFPT) calculated phonon frequencies in CsCuCl$_3$. The corresponding lattice constants are $a = b = 7.19$ Å, $c = 18.08$ Å. The modes were given for the wavevector at the $\Gamma$ point in the Brillouin zone.

| Mode Frequency | Mode Sym (R) | Exp. (300K) (cm$^{-1}$) | Exp. (9.2 K) (cm$^{-1}$) | DFPT (cm$^{-1}$) | Vibrational Assignment |
|---|---|---|---|---|---|
| $\omega_1$ | E$_1$ | | 109 | 105.8 | $\delta_{as}$ (Cl-Cu-Cl) str. |
| $\omega_2$ | E$_2$ | | 115 | 116.1 | $\delta_s$ (Cl-Cu-Cl) str. |
| $\omega_3$ | E$_2$ | 115 | 121 | 122.3 | $\tau$ (Cl-Cu-Cl) wk. [42,45–47] |
| $\omega_4$ | E$_2$ | 132 | 136 | 138.1 | $\omega$ (Cl-Cu-Cl) str. [45–47] |
| $\omega_5$ | E$_2$ | | 145 | 152.6 | $\rho$ (Cl-Cu-Cl) str. |
| $\omega_6$ | E$_1$ | | 155 | 156.8 | $\gamma$ (Cl-Cu-Cl) str. |
| $\omega_7$ | A | 155 | 161 | 160.9 | $\gamma$ (Cl-Cu-Cl) str. [42,45,47] |
| $\omega_8$ | E$_1$ | | 167 | 164.0 | $\omega$ (Cl-Cu-Cl) str. |
| $\omega_9$ | E$_2$ | 177 | 180 | 175.5 | $\tau$ (Cl-Cu-Cl) str. [42,45] |
| $\omega_{10}$ | E$_2$ | | 190 | 191.4 | $\nu$ (Cu-Cl) wk. |
| $\omega_{11}$ | E$_1$ | 191 | 196 | 210.3 | $\nu_s$ (Cl-Cu-Cl) wk. [42,45] |
| $\omega_{12}$ | A | 242 | 245 | 245.9 | $\nu_s$ (Cl-Cu-Cl) str. [42,45–47] |
| $\omega_{13}$ | E$_1$ | 265 | 267 | 266.4 | $\nu_{as}$ (Cl-Cu-Cl) wk. [42,45,47] |
| $\omega_{14}$ | A | | 276 | 267.4 | $\nu_{as}$ (Cl-Cu-Cl) str. |
| $\omega_{15}$ | E$_2$ | 286 | 292 | 273.7 | $\nu_s$ (Cl-Cu-Cl) str. [42,45–47] |

Abbreviations: δ: bending; τ: twisting/torsion; ω: wagging; ρ: rocking; γ: scissoring; ν: stretching; as: asymmetric; s: symmetric; str: strong; wk: weak

## B. Raman spectroscopy at low-temperatures in CsCuCl$_3$ - Spin-phonon coupling

As was previously described, the CsCuCl$_3$ exhibits an antiferromagnetic ordering at $T_N = 10.7$ K produced by DM interaction, which is allowed by the twist of the [Cu$_2$Cl$_9$]$^{5-}$ dimer unit along the [001] direction [13,49]. This magnetic ordering could induce a spin-phonon coupling (SPC) in CsCuCl$_3$. To investigate such coupling, we performed low-temperature Raman spectroscopy. The temperature-dependent Raman spectra of CsCuCl$_3$ in the low-temperature range from room temperature down to 9.2 K are shown in Figure 2. As discussed before, the material does not exhibit a structural phase transition (SPT) within the temperature range investigated. In the absence of a SPT, the behavior of the phonon mode ($\omega$) and the Full Width at Half Maximum (FWHM = $\Gamma$) can be described by the Balkanski Model, which is given by the following equations:

$$\omega(T) = \omega_0 + C\left[1 + \frac{2}{e^x - 1}\right] + D\left[1 + \frac{3}{e^y - 1} + \frac{3}{(e^y - 1)^2}\right] \qquad (2)$$

and

$$\Gamma(T) = A\left[1 + \frac{2}{e^x - 1}\right] + B\left[1 + \frac{3}{e^y - 1} + \frac{3}{(e^y - 1)^2}\right] \qquad (3)$$

where $A$, $B$, $C$, and $D$ are constants referring to the strength of the anharmonic contributions and the $\omega_0$ is the zero-temperature frequency of the corresponding vibrational mode without spin-phonon interaction. To simplify the analysis, the dimensionless parameters $x = \hbar\omega_0/2k_BT$ and $y = \hbar\omega_0/3k_BT$ were also used in these equations. The fit of the experimental data of the Raman mode parameters by equations (2) and (3) (see Figure S3) allows for a detailed description of the paramagnetic phase ($T$ > 15 K) of CsCuCl₃, with the values of the anharmonic contributions for each mode summarized in Table SIII of the Supplementary Material.

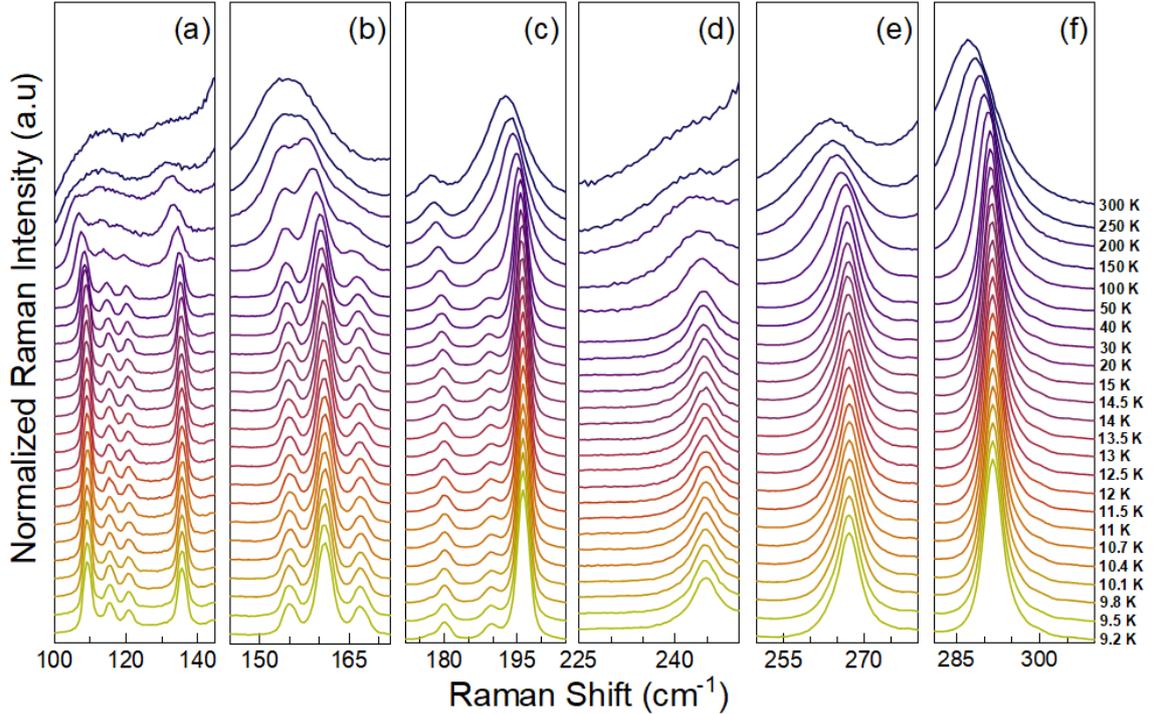

**Figure 2** – Temperature-dependent Raman spectra of CsCuCl₃ in the ranges (a) 100-145 cm⁻¹, (b) 145-175 cm⁻¹, (c) 175-225 cm⁻¹, (d) 225-250 cm⁻¹, (e) 250-280 cm⁻¹, and (f) 280-310 cm⁻¹.

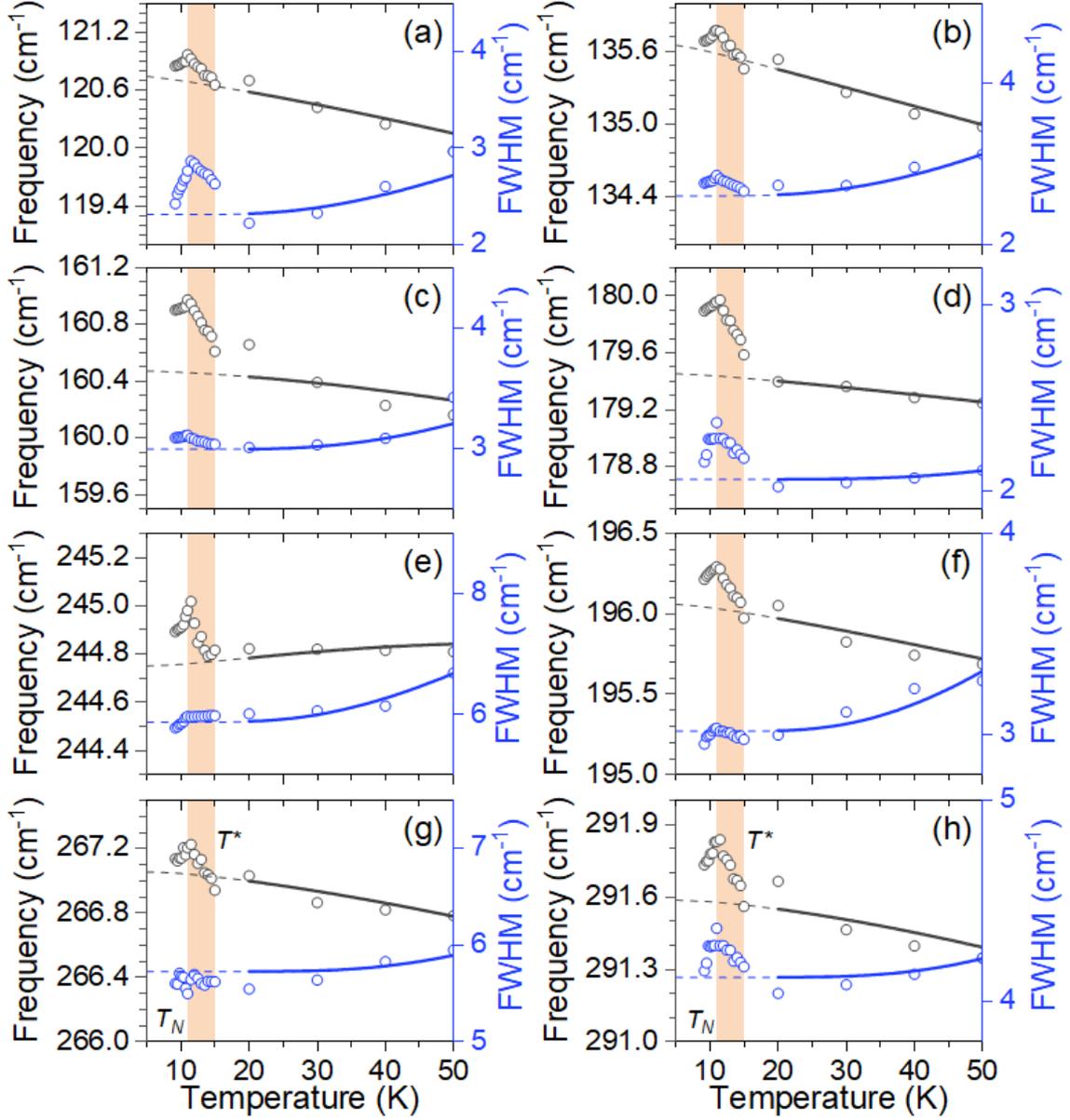

**Figure 3** – Temperature dependence of the phonon frequencies (left axes) and FWHM (right axes) for selected phonons in the paramagnetic and antiferromagnetic phases. Solid curves are the fits using equations (2) and (3). The orange area describes an anomalous hardening region ($T_N = 10.7$ K < $T$ < $T^* = 15$ K) for (a) $\omega_3$, (b) $\omega_4$, (c) $\omega_7$, (d) $\omega_9$ (e) $\omega_{11}$, (f) $\omega_{12}$, (g) $\omega_{13}$, and (h) $\omega_{15}$ phonons.

The phonons depicted in Figure 3 exhibit anomalies in both $\omega(T)$ and $\Gamma(T)$ around $T^* = 15$ K. A sudden phonon hardening is observed at $T^*$ remains until the antiferromagnetic transition temperature $T_N = 10.7$ K. Below $T_N$, the Raman modes exhibit a noticeable softening. Notably, $T^*$ introduces a novel low-temperature correlation within the system unrelated to any unusual lattice distortion, electronic phase transition, or reported structural/magnetic phase transition in the compound. This intriguing outcome strongly suggests the presence of SPC in CsCuCl₃, manifested by the alteration of both mode parameters (wavenumber position and full width at half-maximum, FWHM) through phonon renormalization. While these frequency shifts are typically subtle, often on the order of 1 cm$^{-1}$ or smaller, their systematic nature lends robustness to the findings.

The role of spin-phonon interaction can be elucidated through the static spin-spin correlation average, represented as $\Delta\omega = \omega - \omega_0 = \lambda\langle\boldsymbol{S}_i \cdot \boldsymbol{S}_j\rangle$, where $\lambda$ is the coupling constant, and $\langle\boldsymbol{S}_i \cdot \boldsymbol{S}_j\rangle$ denotes the correlation between the neighboring spins in the $i$ and $j$ sites [50–52]. Consequently, at $T > T^*$ (paramagnetic phase), the $\langle\boldsymbol{S}_i \cdot \boldsymbol{S}_j\rangle$ term is null, due to the absence of the spin ordering. However, in the temperature range $T_N < T < T^*$, a distinct mode frequency renormalization is already discernible prior to $T_N$. This observation indicates unconventional magnetic correlations (frustration or quantum phase, for example) occurring in CsCuCl$_3$ preceding the onset of antiferromagnetic ordering. Analogous behavior has been reported in antiferromagnetic RMn$_2$O$_5$ (R=Bi, Eu, Dy) compounds [53,54].

The magnetic order may couple to the phonon frequencies through modulation of the DM interaction parameter. However, for $T_N < T$, the contribution of the softening is driven by the AFM effect. Therefore, it is convenient to estimate the magnetic contribution to the renormalization of the phonon frequency as a function of $\Delta\omega$. The easiest way to do this is by estimating the $\Delta\omega$ dependence with the mean-field approach $((\langle S^z\rangle)/S)^2$) described by the molecular-field approximations mechanism as $(M(T)/M_0)^2$, where $M(T)$ is the temperature dependence of magnetization, and $M_0$ is the magnetization at zero temperature [55,56].

To obtain the $(M(T)/M_0)^2$ we employed the Yamamoto *et al.* [20] longitudinal susceptibility ($\chi_\parallel$) data at low temperatures obtained by the authors at $H = 1$ T. Thus, based on the spin-phonon coupling mechanism proposed by Granado *et al.* [44], the phonon renormalization induced by the SPC could be reduced as:

$$\Delta\omega = \omega - \omega_0 = \lambda\langle\boldsymbol{S}_i \cdot \boldsymbol{S}_j\rangle \propto \left(\frac{M(T)}{M_0}\right)^2 \tag{4}$$

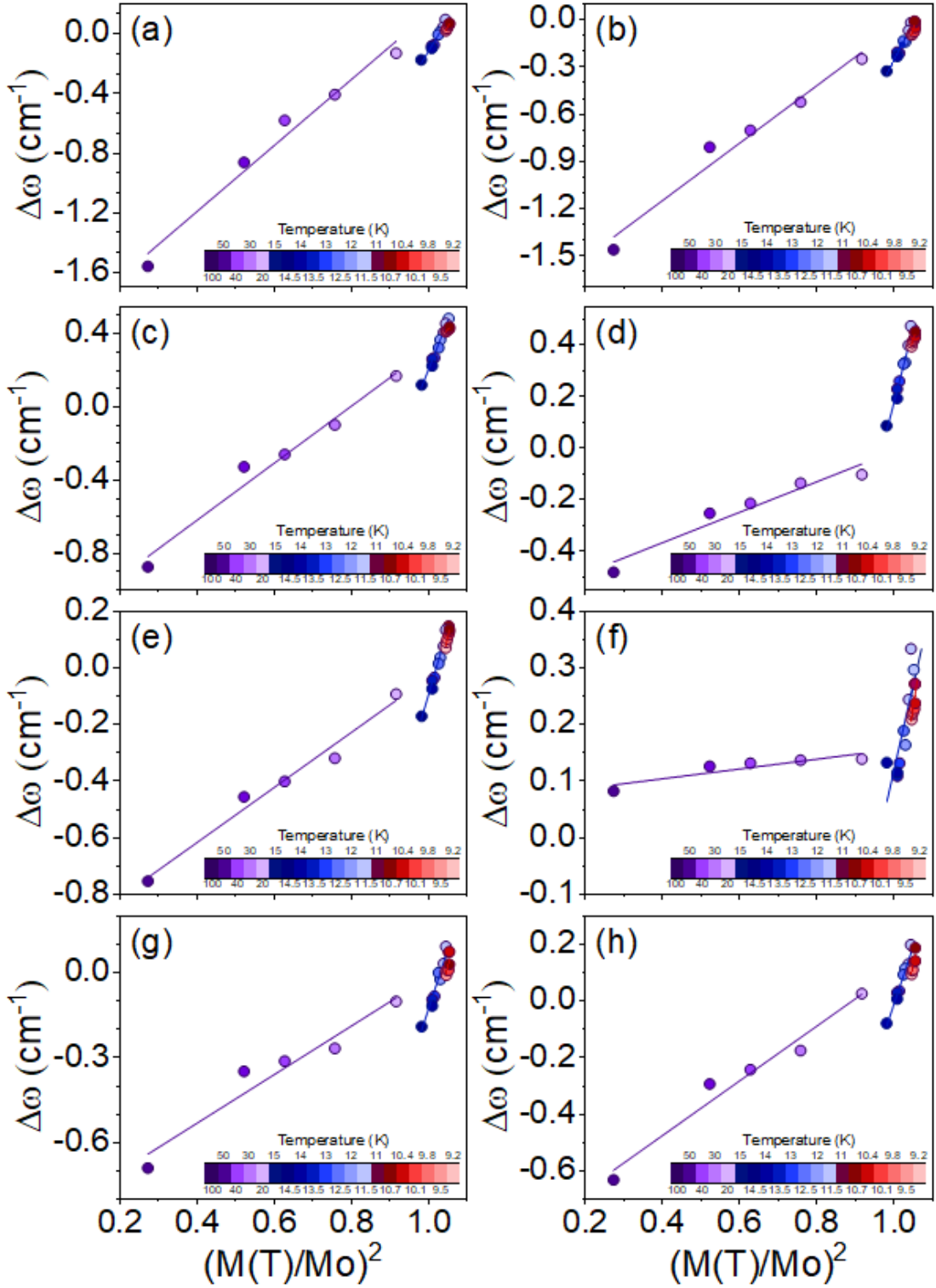

**Figure 4** – Temperature-dependence from the anharmonic behavior of selected phonon as a function of $(M(T)/M_0)^2$ for (a) $\omega_3$, (b) $\omega_4$, (c) $\omega_7$, (d) $\omega_9$ (e) $\omega_{11}$, (f) $\omega_{12}$, (g) $\omega_{13}$, and (h) $\omega_{15}$ phonons. (The purple, blue and red are linear guides for the eyes separated for each region).

Figure 4 shows the $\Delta\omega$ as a function of $(M(T)/M_0)^2$ for the selected phonons. The model, as expressed in equation (4), elucidates a direct linear correspondence between the phonon renormalization phenomenon and the adjustments in the gradient of the linear trend within distinct temperature intervals ($T^* > 15$ K; $T_N < T < T^*$; $T_N < T$). This correspondence firmly establishes the existence of SPC. Notably, the discernible alteration in the phonon behavior at $T^*$ distinctly suggests the onset of SPC within CsCuCl$_3$. The slopes derived from SPC-related are summarized in Table SIII, provided in the Supplementary Material. These slope variations, evident in Figure 4, are consistent with SPC-related materials such as Cu$_2$OCl$_2$ [58], thus robustly substantiating the presence of SPC in CsCuCl$_3$. This congruence bolsters the proposed models that underpin the correction of phonon energy renormalization, intensifying our confidence in the tangible manifestation of SPC within CsCuCl$_3$. As previously mentioned, the detected anomalies (occurring within the temperature range of $T_N < T < T^*$) in the low temperature phonon dependence stand apart from any indications of lattice distortion, electronic phase transitions, or previously reported structural or magnetic phase transitions within the material. This distinct dissociation from these conventional factors suggests a potential antecedent magnetic frustration effect, which appears to be concurrent with the onset of the observed SPC phenomenon.

### C. Pressure-Induced structural phase transition on CsCuCl$_3$

As was previously discussed, CsCuCl$_3$ exhibits pressure-induced quantum phases, which enhances the neighboring spins in the sample and modifies the magnetic diagram by the enhanced Dzyaloshinskii-Moriya interaction [18,19]. Figure 5(a) shows the pressure-dependent Synchrotron powder X-ray diffraction (SPXRD) pattern obtained from CsCuCl$_3$. The diffraction patterns are well described by the same $P6_522$ structure up to 2.38 GPa. Above the critical pressure, $P_c = 3.69$ GPa, the diffractograms exhibit abrupt changes, indicating the presence of new reflections, which can be attributed to a pressure-induced structural phase transition (SPT). At around $P = 2.91$ GPa, both low-pressure and high-pressure phases coexist. Remarkably, the emergence of new diffraction peaks, such as the splitting of the 6.6° band and the appearance of the distribution of the peak around 10° (see Figures 5(b-c)), indicates a decrease in symmetry of the *HP-phase*. No remarkable changes in the diffraction patterns were detected up to 9 GPa, indicating that the CsCuCl$_3$ does not undergo any other phase transition within the maximum pressure range explored in this work.

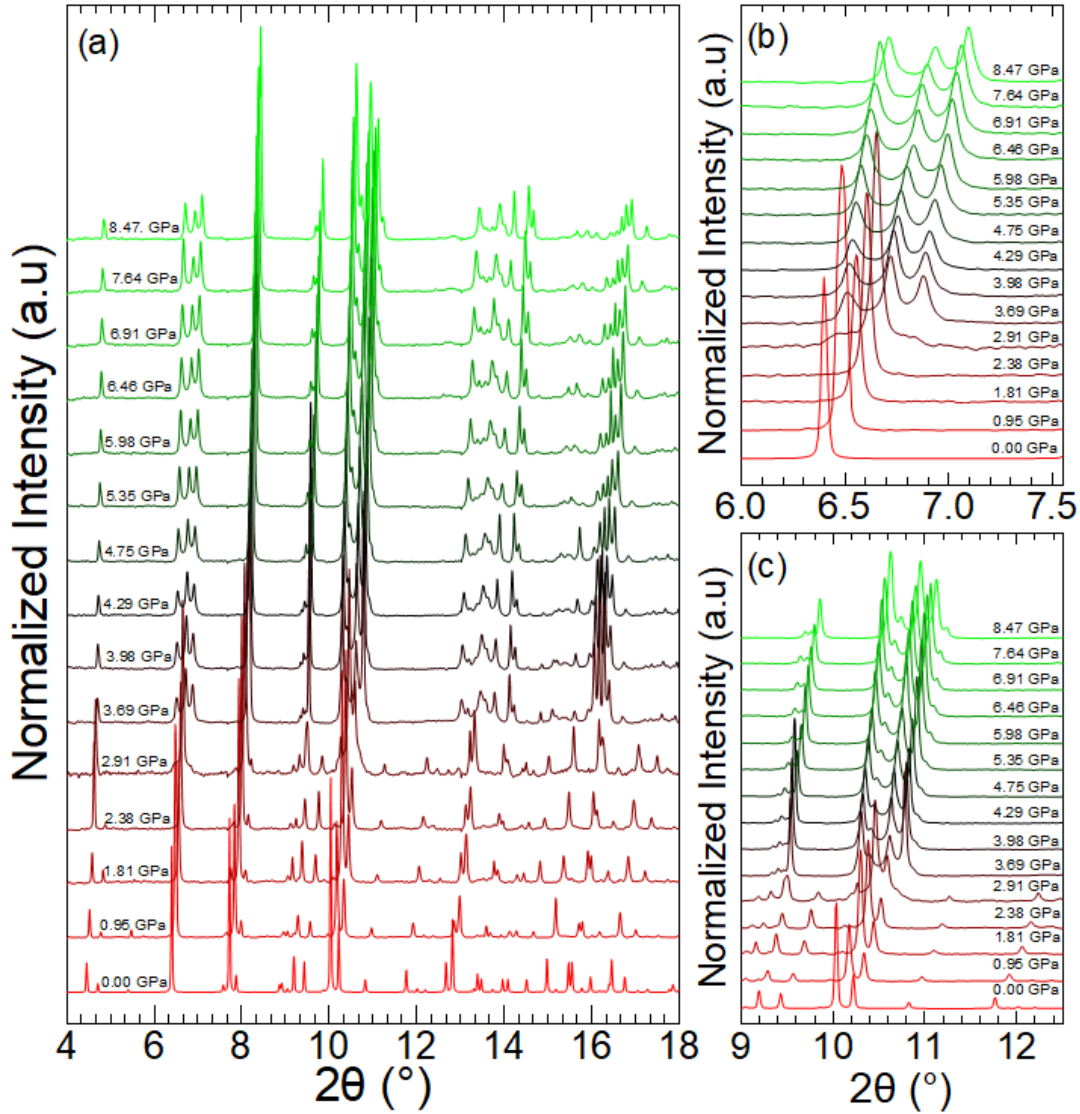

**Figure 5** – (a) Pressure-dependent SPXRD pattern of CsCuCl₃. (b) Pressure-dependent SPXRD pattern zoom around the $2\theta$ (6-7.6 range). (c) Pressure-dependent SPXRD pattern zoom around the $2\theta$ (9-13 range).

All patterns of the low-pressure phase were refined using the Rietveld method implemented in the EXPO2014 [37] software with the hexagonal $P6_522$ structure obtained from our single-crystal X-ray diffraction measurement. To investigate the crystal structure of the high-pressure phase, primarily the X-ray powder pattern was compared with the one corresponding to the orthorhombic CsCuBr₃ ($C222_1$) structure reported in [59] (ICSD: 10184) as a possible solution. However, the reflections of the simulated DRX patterns did not match the experimental data. Thus, our results differ from the high-temperature SPT [15,60,61], which involves a hexagonal-to-hexagonal $P6_122$ or $P6_522 \rightarrow P6_3/mmc$ at 423 K, and the high-pressure hexagonal sub-cell ($P6_3/mmc$) encountered at 3.04 GPa by Christy *et al.* [62] (ISCD:32503). A comparison of all diffractograms is provided in Figure 6.

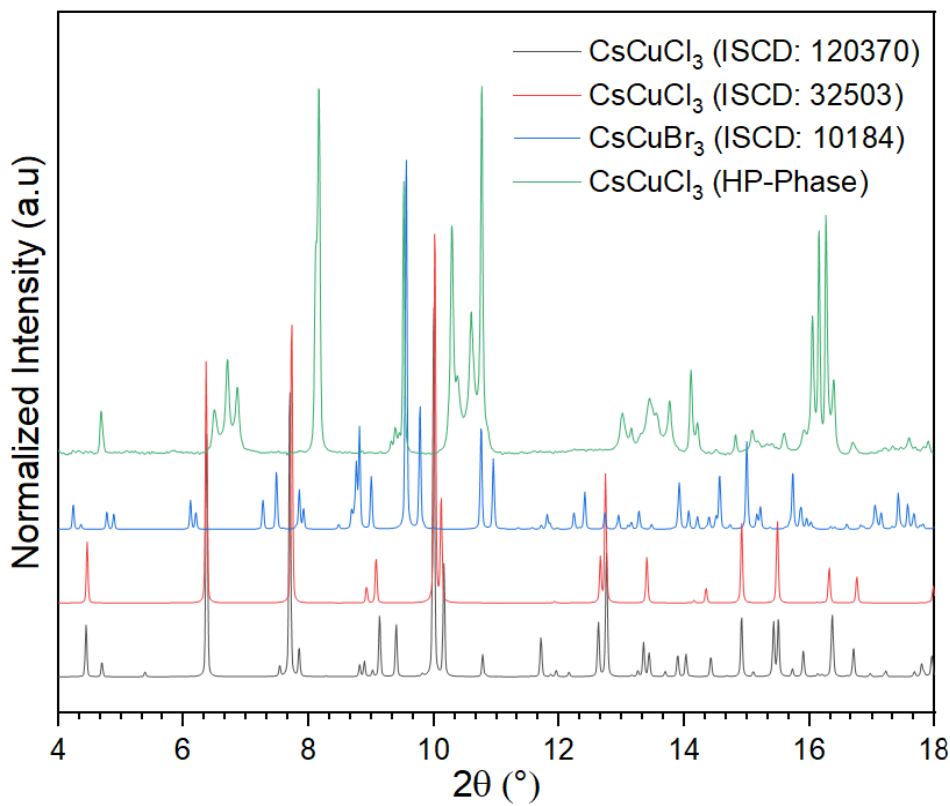

**Figure 6** – Simulated diffraction patterns of the CsCuCl$_3$ (*P*6$_5$22 - ICSD: 120370), CsCuCl$_3$ (*P*6$_3$/*mmc* - ISCD:32503), CsCuBr$_3$ (*C*222$_1$ - ICSD: 10184) and the CsCuCl$_3$ synchrotron powder X-ray diffraction pattern obtained at 3.63 GPa.

The high-pressure phase was refined by the *Le-Bail* method using the EXPO2014 [37] software. The results suggest a base-centered monoclinic structure (Type C). Unfortunately, the low diffraction intensity due to the sample texture and preferential orientation does not allow a high-quality number of peaks, which makes the diffractogram refinement challenging. Since at room pressure, CsCuCl$_3$ adopts the *P*6$_5$22 (*a* = *b* = 7.2168 (10) Å, *c* = 18.1853 (5) Å) with a *3*-fold rotational symmetry along the principal *c*-axis. The cooper (located at the center) and chlorine (vertices of the octahedron) atoms form a distorted octahedron coordination geometry [CuCl$_6$]$^{4-}$. Each octahedron shares a common edge, forming a linear chain of octahedra that runs through the crystal structure along the principal axis. Cs$^+$ ions occupy the interstitial sites between the displayed chains.

Under high pressure, the structure undergoes the structural phase transition to the monoclinic one ($a$ = 6.8875 (12) Å, $b$ = 6.7918 (2) Å, $c$ = 5.8539 (10) Å, β = 93.76 (4) °). To facilitate the discussion of the relationship between both phases, a transformation was made on the crystal cell parameters: $a_h = a_m; \sqrt{3}b_h = b_m; c_h = 3c_m.$ (see Figure S4 in the Supplementary Material). Figure 7 shows the pressure dependence of the reduced unit-cell parameters; a shift over all lattice parameters was observed, indicating crystal modification. The significant discontinuity indicates a first-order transition character.

The resulting structural distortion induced by the SPT can generate uniaxial stress as a function of the compound elastic anisotropy, which can be predicted by the equation of state (EOS) [68–70]. The third-order Birch-Murnaghan EOS was used to fit the pressure-dependent of the unit cell volume (see Fig. 8(f)). The equation was expressed in terms of the volume at zero pressure ($V_0$), the bulk modulus $B_0 = (-V\partial P/\partial V)_T$, and the dimensionless pressure derivative $B' = (\partial B/\partial P)_T$ (dimensionless), which describes as $B_0$ change with pressure [71]:

$$P = \frac{3}{2}B_0\left[\left(\frac{V_0}{V}\right)^{7/3} - \left(\frac{V_0}{V}\right)^{5/3}\right]\left[1 + \frac{3}{4}(B'-4)\left\{\left(\frac{V_0}{V}\right)^{2/3} - 1\right\}\right] \qquad (5)$$

Table II shows the parameters obtained by the pressure-dependent unit cell volume fit using the third-order Birch-Murnaghan EOS. The SPT was accompanied by increased unit-cell volume, which is consistent and expected for high-pressure phases with higher bulk modulus due to their denser and less compressible crystal structure [72]. The $B_0$ values typically range from 10 to 70 GPa for hybrids/inorganic materials based on metal halide perovskites [73–75]. The low bulk modulus of metal halide perovskites is thought to contribute to their unique properties, such as self-healing, ion migration, and low thermal conductivity, which suggest applications in the flexible electronics industry for their ductility [76,77]. The value of $B'$ indicates a slow stiffening of the material, which can be attributed to the first-order structural phase transition, where both low- and high-pressure phases can be related to a unique basic set. The presence of dynamic instability in the sample was not observed, and the positive value of $B'$ compensated for any instability.

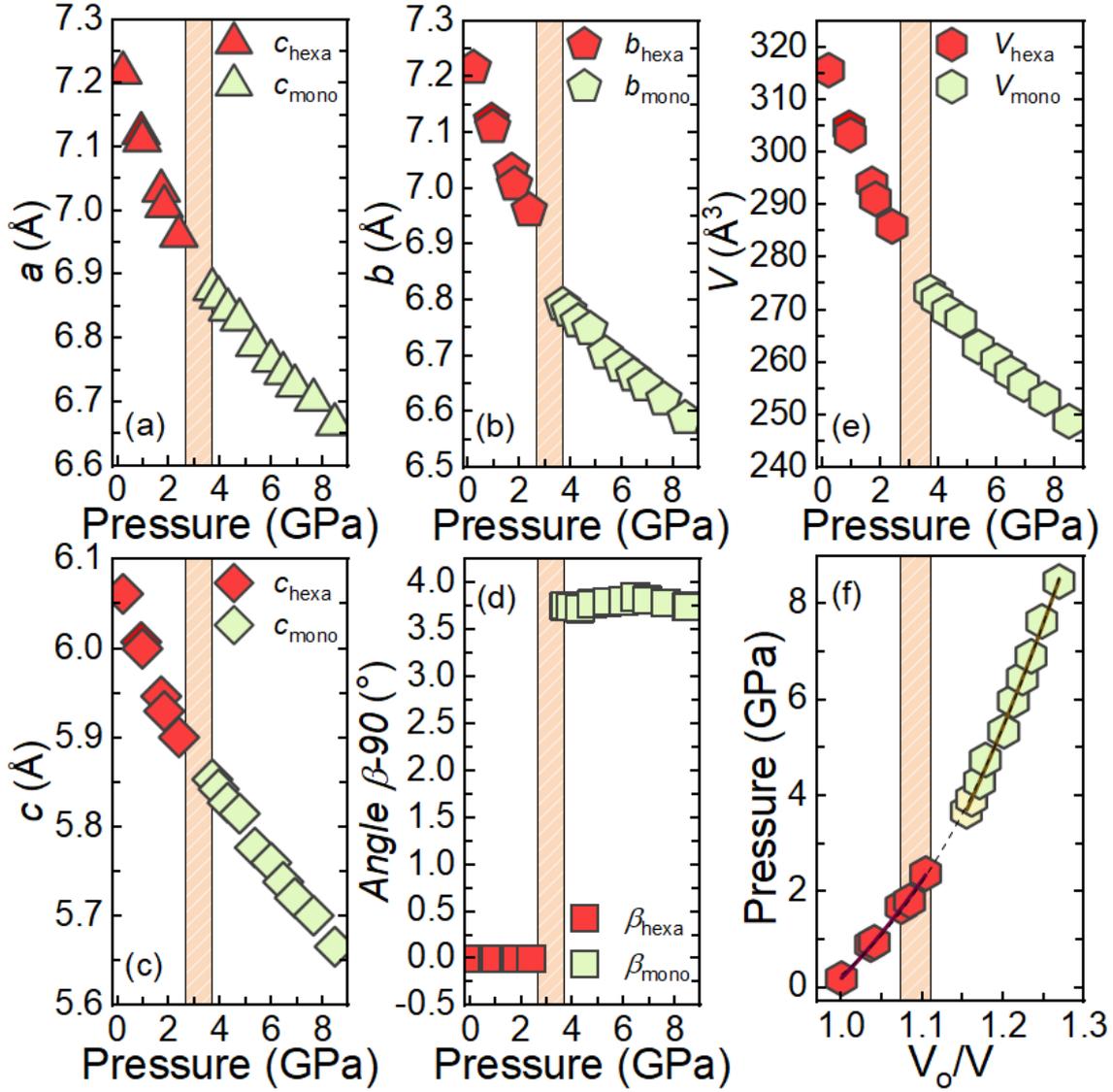

**Figure 7** – (a)-(e) The lattice parameters as a function of pressure. (f) Third-order Birch-Murnaghan fit for each pressure phase. The orange region dashboard stands for the coexisting phases of both crystal structures.

**Table II**. Experimental coefficients of the Murnaghan equation of state for CsCuCl$_3$ in each structural phase.

| Crystal System | P$_{range}$ (GPa) | V$_0$ (Å$^3$) | B$_0$ (GPa) | B' |
|---|---|---|---|---|
| Hexagonal | 0.10 − 2.38 | 319.42 ± 0.02 | 17 ± 3 | 4.39 ± 0.04 |
| Monoclinic | 3.69 − 8.47 | 304.62 ± 0.06 | 27 ± 9 | 4.26 ± 0.07 |

Since Raman spectra are sensitive to crystalline structure, we performed Raman spectroscopy in CsCuCl$_3$ at high-pressure in order to gain insights into Raman-active modes and structural changes in CsCuCl$_3$ across different phases of the material correlated to SPXRD. Figure 8 shows the pressure-dependent Raman spectra of CsCuCl$_3$ up to 7 GPa. From room pressure up to 2.51 GPa, we can observe that the spectra keep the same Raman bands profile. Hence, a reorganization process involving the dimer unit $[Cu_2Cl_9]^{5-}$ was observed in a 2.79-3.57 GPa range, where both *LP*- and *HP*- phases coexist, which means hexagonal and monoclinic phases. Consistent with SPXRD, a sudden change in the Raman spectra pressure-dependence modification is observed at 4.01 GPa, where new phonons are displayed in the Raman spectra, pertinent to a new crystalline phase. Thus, our Raman results confirm those obtained by diffraction.

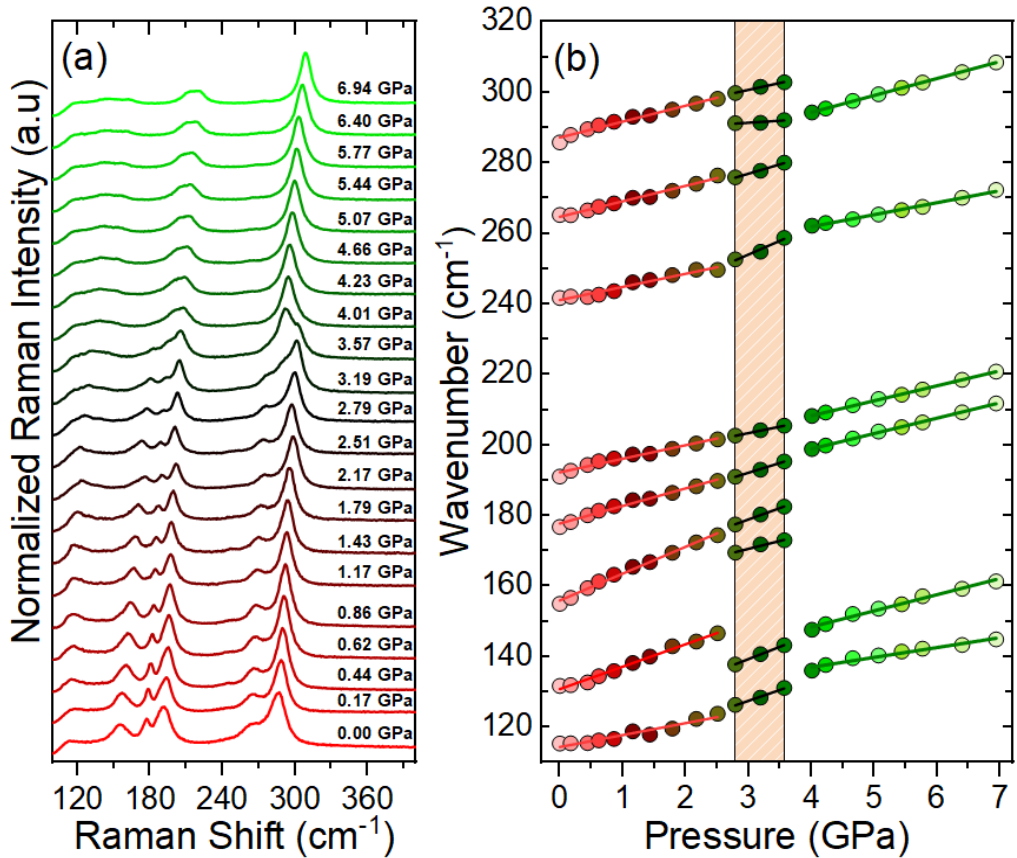

**Figure 8** – (a) Pressure-dependent Raman spectra of $CsCuCl_3$. (b) Pressure dependence of the Raman mode positions. The orange region dashboard represents the coexisting LP- and HP-phase regions.

Note that pressure dependence of the phonons progressively becomes less resolved, probably due to stress induced along the structural phase transition or could be a consequence of the increase in the sample bulk modulus. In general, the pressure dependence of the mode frequencies is linear. We observe this behavior in both phases. Furthermore, the Cl-Cu-Cl bending under pressure contributes more to the reorganization around the structural phase transition for the number of displayed bands. However, the most intense band in the Raman spectra, the Cl-Cu-Cl stretching, is associated with the stronger distortion of $[Cu_2Cl_9]^{5-}$ for the significant redshift. The Grüneisen parameters of each mode (see Table III), which are given by:

$$\gamma_{iT} = (B_0/\omega_i)(\partial\omega_i/\partial P)_T \qquad (6)$$

where $B_0$ is the bulk modulus of each phase, the $(\partial\omega_i/\partial P)_T$ is the proper linear slope and $\omega_i$ is the Raman frequency of the specific Grüneisen parameter ($\gamma_{iT}$). As it was predicted, the Cl-Cu-Cl bending modes have higher values of $\gamma_{iT}$ than the Cu-Cl stretching, contributing to the modification in $[Cu_2Cl_9]^{5-}$. Comparing the phonon frequencies, the Cl-Cu-Cl stretching exhibits more significant shifts, consistent with previous predictions. This result establishes the structural correlations between bulk modulus and the corresponding local Grüneisen parameter. Confirming the XRD results, the Raman spectra show that for pressures up to 9 GPa, the CsCuCl3 material does not undergo amorphization.

**Table III**. Experimental Raman modes for CsCuCl₃ in each structural phase and their corresponding Grüneisen parameters. The material bulk modulus was $B_0 = 17$ GPa and $B_0 = 27$ GPa for each phase, obtained from the fit of the SPXRD pattern by the third-order Birch-Murnaghan equation.

| LP-Phase | | | Coexistence Phase | | | HP-Phase | | |
|---|---|---|---|---|---|---|---|---|
| $\omega_i$ (cm⁻¹) | $(\partial\omega_i/\partial P)_T$ (cm⁻¹/GPa) | $\gamma_{iT}$ | $\omega_i$ (cm⁻¹) | $(\partial\omega_i/\partial P)_T$ (cm⁻¹/GPa) | $\gamma_{iT}$ | $\omega_i$ (cm⁻¹) | $(\partial\omega_i/\partial P)_T$ (cm⁻¹/GPa) | $\gamma_{iT}$ |
| – | – | – | 109 | 6.2 ± 0.6 | – | 126 | 2.8 ± 0.2 | 0.600 |
| 114 | 3.4 ± 0.3 | 0.507 | 118 | 7.0 ± 0.2 | – | 130 | 4.6 ± 0.2 | 0.955 |
| 131 | 6.3 ± 0.2 | 0.818 | 157 | 4.5 ± 0.8 | – | – | – | – |
| 156 | 7.6 ± 0.2 | 0.828 | 160 | 6.5 ± 0.4 | – | – | – | – |
| 178 | 5.0 ± 0.2 | 0.478 | 176 | 5.5 ± 0.3 | – | 182 | 4.4 ± 0.1 | 0.653 |
| 192 | 3.8 ± 0.2 | 0.336 | 193 | 3.6 ± 0.1 | – | 191 | 4.2 ± 0.1 | 0.594 |
| 241 | 3.7 ± 0.3 | 0.261 | 231 | 7.8 ± 0.9 | – | 248 | 3.4 ± 0.1 | 0.370 |
| 265 | 4.4 ± 0.2 | 0.282 | 261 | 5.3 ± 0.4 | – | – | – | – |
| – | – | – | 288 | 1.1 ± 0.5 | – | 275 | 4.8 ± 0.1 | 0.471 |
| 287 | 4.5 ± 0.3 | 0.267 | 289 | 3.9 ± 0.3 | – | – | – | – |

Finally, this transition is reversible since the hexagonal phase was obtained again after the full release of pressure CsCuCl₃ (see Figure S5 of the Supplementary Material). However, it is important to point out that the critical pressure for the decompression cycle was lower than that obtained at the increasing process. The coexisting region could be responsible for the slow liberation of pressure and the readjustment to the original crystal structure. However, the recovery of the low-pressure phase is well preserved. This is expected since the SPT is a first-order one.

# IV. CONCLUSIONS

Investigating the structural phase transitions and their related physical phenomena in halide perovskite-like materials under extreme conditions is significant for understanding their fundamental properties and exploring their potential applications in various fields. In the case of $CsCuCl_3$, applying low-temperature Raman spectroscopy has unveiled a distinct spin-phonon coupling mechanism. This coupling is evidenced by discernible shifts and broadening in the frequencies and FWHM of select phonons linked to Cl-Cu-Cl bending and stretching modes. This spin-phonon coupling generates an expected contribution to the phonon frequencies, observed at $T^* = 15$ K, suggesting a magnetic frustration within the material. This correlation between spin and phonon behaviors becomes more apparent with a discernible hardening of phonon temperature dependence around the anticipated renormalization temperature $T_N = 10.7$ K, reflective of the underlying antiferromagnetic effects. Utilizing the $\Delta\omega$ as a function of $(M(T)/M_0)^2$ reinforces the discernible transformation linked to the spin-phonon coupling renormalization process at $T^*$.

Conversely, conducting high-pressure investigations involving synchrotron powder X-ray diffraction and Raman spectroscopy on $CsCuCl_3$ has brought a first-order Structural Phase Transition (SPT) at the critical pressure $P_c = 3.69$ GPa from the hexagonal $P6_522$ space group to the base-centered monoclinic type $C$ cell. An intriguing feature emerges wherein the direct correlation in crystal axes engenders a gradual and natural crystal reconfiguration devoid of significant dynamic instability. This observation implies a degree of inherent flexibility within $CsCuCl_3$. Notably, the coexistence of phases is discernible within the pressure range from 2.79 to 3.57 GPa. The SPT involves the reorganization of the internal $[Cu_2Cl_9]^{5-}$ dimer unit, with Cl-Cu-Cl bending contributing more than Cl-Cu-Cl stretching. The prominent shift presence and highest $\gamma_{iT}$ value associated with the displayed band indicates heightened distortion. Moreover, our analysis underscores the reversible nature of the SPT; however, residual strain pressure influences the modification of the $P_c$ value upon pressure decrease. These findings supplied additional information about modifying the $[Cu_2Cl_9]^{5-}$ under pressure and enhanced our understanding of the lattice distortion at external pressure.

**Acknowledgments**

This investigation was financed in part by the Fundação Cearense de Apoio ao Desenvolvimento Científico e Tecnológico (FUNCAP/PRONEX PR2-0101-00006.01.00/15), the Conselho Nacional de Pesquisa e Desenvolvimento do Brasil (CNPq – Projeto 140390/2019-7), Coordenação de Aperfeiçoamento de Pessoal de Nível Superior do Brasil (CAPES) – Finance Code 001 (CAPES – Projeto 88882.349932/2019-01), Fundação de Apoio à Pesquisa do Estado de São Paulo (FAPESP, Projeto:2018/00823-0). This research used facilities of the Brazilian Synchrotron Light Laboratory (SIRIUS-LNLS), part of the Brazilian Center for Research in Energy and Materials (CNPEM), a private non-profit organization under the supervision of the Brazilian Ministry for Science, Technology, and Innovations (MCTI). The EMA beamline and the LCTE staff are acknowledged for their assistance during the experiments (Proposal 20210152). All calculations were conducted using the computational resources of the Centro Nacional de Processamento de Alto Desempenho in São Paulo (CENAPAD-SP) Project 823.

# V. REFERENCES

[1]   R. Roccanova, A. Yangui, G. Seo, T. D. Creason, Y. Wu, D. Y. Kim, M. H. Du, and B. Saparov, *Bright Luminescence from Nontoxic CsCu₂X₃ (X = Cl, Br, I)*, ACS Mater Lett **1**, 459 (2019).

[2]   F. Cao and L. Li, *Progress of Lead-Free Halide Perovskites: From Material Synthesis to Photodetector Application*, Adv Funct Mater **31**, (2021).

[3]   X. Zhang, B. Zhou, X. Chen, and W. W. Yu, *Reversible Transformation between Cs₃Cu₂I₅ and CsCu2I₃ Perovskite Derivatives and Its Anticounterfeiting Application*, Inorg Chem **61**, 399 (2022).

[4]   Y. Lu, G. Li, S. Fu, S. Fang, and L. Li, *CsCu₂I₃ Nanocrystals: Growth and Structural Evolution for Tunable Light Emission*, ACS Omega **6**, 544 (2021).

[5]   M. ben Bechir and M. H. Dhaou, *Study of Charge Transfer Mechanism and Dielectric Relaxation of CsCuCl₃ Perovskite Nanoparticles*, Mater Res Bull **144**, (2021).

[6]   Y. Zheng, X. Yuan, J. Yang, Q. Li, X. Yang, Y. Fan, H. Li, H. Liu, and J. Zhao, *Cu Doping-Enhanced Emission Efficiency of Mn²⁺ in Cesium Lead Halide Perovskite Nanocrystals for Efficient White Light-Emitting Diodes*, J Lumin **227**, (2020).

[7]   R. Wu, Z. Bai, J. Jiang, H. Yao, and S. Qin, *Research on the Photoluminescence Properties of Cu²⁺-Doped Perovskite CsPbCl₃ quantum Dots*, RSC Adv **11**, 8430 (2021).

[8]   L. T. Nguyen and R. J. Cava, *Hexagonal Perovskites as Quantum Materials*, Chem Rev **121**, 2935 (2021).

[9]   S. Fop, K. S. McCombie, E. J. Wildman, J. M. S. Skakle, and A. C. Mclaughlin, *Hexagonal Perovskite Derivatives: A New Direction in the Design of Oxide Ion Conducting Materials*, Chemical Communications **55**, 2127 (2019).

[10]  L. Balents, *Spin Liquids in Frustrated Magnets*, Nature **464**, 199 (2010).

[11]  J. Collocott and J. A. Rayne, *Low-Temperature Heat Capacity of Linearachain Magnetic Compounds CsNiCl₃, RbNiCl₃, and CsCuCl₃*, J Appl Phys **61**, 4404 (1987).

[12]  A. Miyake, J. Shibuya, M. Akaki, H. Tanaka, and M. Tokunaga, *Magnetic Field Induced Polar Phase in the Chiral Magnet CsCuCl₃*, Phys Rev B **92**, (2015).

[13]  H. Ueda, E. Skoropata, M. Burian, V. Ukleev, G. S. Perren, L. Leroy, J. Zaccaro, and U. Staub, *Conical Spin Order with Chiral Quadrupole Helix in CsCuCl₃*, Phys Rev B **105**, (2022).

[14]  V. P. Plakhty, J. Wosnitza, N. Martin, Y. Marchi, O. P. Smirnov, B. Grenier, and S. V. Gavrilov, *Isostructural Transition Coupled with Spin Ordering in CsCuCl₃: A Spatially Frustrated Spiral Crystal Lattice*, Phys Rev B **79**, (2009).

[15] C. J. Kroese and W. J. A. Maaskant, *The Relation between the High-Temperature and Room-Temperature Structure of CsCuCl₃*, Chem Phys 224 (1974).

[16] S. Cui, Y. Chen, S. Tao, J. Cui, C. Yuan, N. Yu, H. Zhou, J. Yin, and X. Zhang, *Synthesis, Crystal Structure and Photoelectric Response of All-Inorganic Copper Halide Salts CsCuCl₃*, Eur J Inorg Chem **2020**, 2165 (2020).

[17] K. Nihongi, T. Kida, Y. Narumi, J. Zaccaro, Y. Kousaka, K. Inoue, K. Kindo, Y. Uwatoko, and M. Hagiwara, *Magnetic Field and Pressure Phase Diagrams of the Triangular-Lattice Antiferromagnet CsCuCl₃ Explored via Magnetic Susceptibility Measurements with a Proximity-Detector Oscillator*, Phys Rev B **105**, (2022).

[18] A. Sera, Y. Kousaka, J. Akimitsu, M. Sera, and K. Inoue, *Pressure-Induced Quantum Phase Transitions in the S= 1/2 Triangular Lattice Antiferromagnet CsCuCl₃*, Phys Rev B **96**, (2017).

[19] M. Hosoi, H. Matsuura, and M. Ogata, *New Magnetic Phases in the Chiral Magnet CsCuCl₃ under High Pressures*, J Physical Soc Japan **87**, (2018).

[20] D. Yamamoto, T. Sakurai, R. Okuto, S. Okubo, H. Ohta, H. Tanaka, and Y. Uwatoko, *Continuous Control of Classical-Quantum Crossover by External High Pressure in the Coupled Chain Compound CsCuCl₃*, Nat Commun **12**, (2021).

[21] Y. Liu, Q. Liu, Y. Liu, X. Jiang, X. Zhang, and J. Zhao, *Effects of Spin-Phonon Coupling on Two-Dimensional Ferromagnetic Semiconductors: A Case Study of Iron and Ruthenium Trihalides*, Nanoscale **13**, 7714 (2021).

[22] B. H. Zhang, Y. S. Hou, Z. Wang, and R. Q. Wu, *First-Principles Studies of Spin-Phonon Coupling in Monolayer Cr₂Ge₂Te₆*, Phys Rev B **100**, (2019).

[23] G. Qin, H. Wang, L. Zhang, Z. Qin, and M. Hu, *Giant Effect of Spin-Lattice Coupling on the Thermal Transport in Two-Dimensional Ferromagnetic CrI₃*, J Mater Chem C Mater **8**, 3520 (2020).

[24] D. Pesin and L. Balents, *Mott Physics and Band Topology in Materials with Strong Spin-Orbit Interaction*, Nat Phys **6**, 376 (2010).

[25] C. H. Sohn et al., *Strong Spin-Phonon Coupling Mediated by Single Ion Anisotropy in the All-In-All-Out Pyrochlore Magnet Cd₂Os₂O₇*, Phys Rev Lett **118**, (2017).

[26] J. Son, B. C. Park, C. H. Kim, H. Cho, S. Y. Kim, L. J. Sandilands, C. Sohn, J. G. Park, S. J. Moon, and T. W. Noh, *Unconventional Spin-Phonon Coupling via the Dzyaloshinskii–Moriya Interaction*, NPJ Quantum Mater **4**, (2019).

[27] Bruker (2018), *APEX3*, Bruker AXS Inc.

[28] Bruker (2018), *SAINT*, Bruker AXS Inc.

[29] G. M. Sheldrick, *SHELXT – Integrated Space-Group and Crystal- Structure Determination Research Papers*, Acta Crystallogr A Found Adv 3 (2015).

[30] G. M. Sheldrick, *Crystal Structure Refinement with SHELXL*, Acta Crystallogr C Struct Chem 3 (2015).

[31] O. V Dolomanov, L. J. Bourhis, R. J. Gildea, J. A. K. Howard, and H. Puschmann, *OLEX2: A Complete Structure Solution, Refinement and Analysis Program*, J Appl Crystallogr 2008 (2009).

[32] C. F. Macrae, I. Sovago, S. J. Cottrell, P. T. A. Galek, E. Pidcock, M. Platings, G. P. Shields, J. S. Stevens, M. Towler, and P. A. Wood, *Mercury 4.0: From Visualization to Analysis, Design and Prediction*, J Appl Crystallogr 226 (2020).

[33] K. Momma and F. Izumi, *VESTA 3 for Three-Dimensional Visualization of Crystal, Volumetric and Morphology Data*, J Appl Crystallogr 1272 (2011).

[34] G. Shen et al., *Toward an International Practical Pressure Scale: A Proposal for an IPPS Ruby Gauge (IPPS-Ruby2020)*, High Press Res 299 (2020).

[35] M. Wojdyr, *Fityk: A General-Purpose Peak Fitting Program*, J Appl Crystallogr **43**, 1126 (2010).

[36] C. Prescher and V. B. Prakapenka, *DIOPTAS: A Program for Reduction of Two-Dimensional X-Ray Diffraction Data and Data Exploration*, High Press Res **35**, 223 (2015).

[37] A. Altomare, C. Cuocci, C. Giacovazzo, A. Moliterni, R. Rizzi, N. Corriero, and A. Falcicchio, EXPO2013*: A Kit of Tools for Phasing Crystal Structures from Powder Data*, J Appl Crystallogr **46**, 1231 (2013).

[38] P. Giannozzi et al., *QUANTUM ESPRESSO: A Modular and Open-Source Software Project for Quantum Simulations of Materials*, Journal of Physics Condensed Matter **21**, (2009).

[39] P. Giannozzi et al., *Advanced Capabilities for Materials Modelling with Quantum ESPRESSO*, Journal of Physics Condensed Matter **29**, (2017).

[40] D. R. Hamann, *Optimized Norm-Conserving Vanderbilt Pseudopotentials*, Phys Rev B Condens Matter Mater Phys **88**, (2013).

[41] J. P. Perdew, K. Burke, and M. Ernzerhof, *Generalized Gradient Approximation Made Simple*, Phys Rev Lett **77**, (1996).

[42] J. Petzelt, I. Gregora, V. Vorliiiek, J. Fousek, B. Biezina, G. V. Kozlov, and A. A. Volkov, *Far-Infrared and Raman Spectroscopy of the Phase Transition in $CsCuCl_3$*, Journal of Raman Spectroscopy **10**, 187 (1981).

[43] G. Mattney Cole, Jr. Charles F. Putnik, and Smith L. Holt, *Physical Properties of Linear-Chain Systems. II. Optical Spectrum of $CsMnBr_3$*, Inorg Chem **14**, 2219 (1975).

[44] U. Kambli and H. U. Giidel, *Optical Absorption and Luminescence Studies of Antiferromagnetic $RbMnCl_3$ and $CsMnCl_3$*, Journal of Physics C: Solid State Physics **17**, 4041 (1984).

[45] C. W. Tomblin, G. D. Jones, and R. W. G. Syme, *Raman Scattering and Infrared Absorption Spectra of $Co^{2+}$ Ions in $CsMgBr_3$ and $CsCdBr_3$*, J. Phys. C: Solid State Phys **17**, 4345 (1984).

[46] E. Jara, J. A. Barreda-Argüeso, J. González, F. Rodríguez, and R. Valiente, *Origin of the Piezochromism in $Cs_2CuCl_4$: Electron-Phonon and Crystal-Structure Correlations*, Phys Rev B **99**, (2019).

[47] L. Nataf, F. Aguado, I. Hernández, R. Valiente, J. González, M. N. Sanz-Ortiz, H. Wilhelm, A. P. Jephcoat, F. Baudelet, and F. Rodríguez, *Volume and Pressure Dependences of the Electronic, Vibrational, and Crystal Structures of $Cs_2CoCl_4$: Identification of a Pressure-Induced Piezochromic Phase at High Pressure*, Phys Rev B **95**, (2017).

[48] Y. Rodríguez-Lazcano, L. Nataf, and F. Rodríguez, *Electronic Structure and Luminescence of $[(CH_3)_4N]_2MnX_4$ (X=Cl,Br) Crystals at High Pressures by Time-Resolved Spectroscopy: Pressure Effects on the Mn-Mn Exchange Coupling*, Phys Rev B **085115**, 1 (2009).

[49] N. Nakagawa et al., *Magneto-Chiral Dichroism of $CsCuCl_3$*, Phys Rev B **96**, (2017).

[50] L. Du et al., *Lattice Dynamics, Phonon Chirality, and Spin–Phonon Coupling in 2D Itinerant Ferromagnet $Fe_3GeTe_2$*, Adv Funct Mater **29**, (2019).

[51] E. Aytan, B. Debnath, F. Kargar, Y. Barlas, M. M. Lacerda, J. X. Li, R. K. Lake, J. Shi, and A. A. Balandin, *Spin-Phonon Coupling in Antiferromagnetic Nickel Oxide*, Appl Phys Lett **111**, (2017).

[52] R. X. Silva, H. Reichlova, X. Marti, D. A. B. Barbosa, M. W. Lufaso, B. S. Araujo, A. P. Ayala, and C. W. A. Paschoal, *Spin-Phonon Coupling in $Gd(Co_{1/2}Mn_{1/2})O_3$ Perovskite*, J Appl Phys **114**, (2013).

[53] A. F. García-Flores, E. Granado, H. Martinho, C. Rettori, E. I. Golovenchits, V. A. Sanina, S. B. Oseroff, S. Park, and S. W. Cheong, *Magnetically Frustrated Behavior in Multiferroics $RMn_2O_5$ (R=Bi, Eu, and Dy): A Raman Scattering Study*, J Appl Phys **101**, (2007).

[54] A. F. García-Flores, E. Granado, H. Martinho, R. R. Urbano, C. Rettori, E. I. Golovenchits, V. A. Sanina, S. B. Oseroff, S. Park, and S. W. Cheong, *Anomalous Phonon Shifts in the Paramagnetic Phase of Multiferroic $RMn_2O_5$ (R=Bi, Eu, Dy): Possible Manifestations of Unconventional Magnetic Correlations*, Phys Rev B **73**, (2006).

[55] M. A. Prosnikov, A. N. Smirnov, V. Y. Davydov, R. V. Pisarev, N. A. Lyubochko, and S. N. Barilo, *Magnetic Dynamics and Spin-Phonon Coupling in the Antiferromagnet $Ni_2NbBO_6$*, Phys Rev B **98**, (2018).

[56] B. Araújo et al., *Spin-Phonon Coupling in Monoclinic $BiCrO_3$*, J Appl Phys **2020**, 114102 (2020).

[57] E. Granado, A. Garcí, J. A. Sanjurjo, C. Rettori, I. Torriani, F. Prado, R. D. Sá Nchez, A. Caneiro, and S. B. Oseroff, *Magnetic Ordering Effects in the Raman Spectra of La₁₋ₓMn₁₋ₓO₃*, Phys Rev B **60**, 11879 (1999).

[58] B. S. Araújo, A. M. Arévalo-López, J. P. Attfield, C. W. A. Paschoal, and A. P. Ayala, *Spin-Phonon Coupling in Melanothallite Cu₂OCl₂*, Appl Phys Lett **113**, (2018).

[59] T.-I. Li and G. D. Stucky, *Exchange Interactions in Polynuclear Transition Metal Complexes. Structural Properties of Cesium Tribromocuprate(II), CsCuBr₃, a Strongly Coupled Copper(II) System*, Inorg Chem **12**, 441 (1973).

[60] C. J. Kroese, W. J. A. Maaskant, and G.C. Verschoor, *The High-Temperature Structure of CsCuCl₃*, Acta Crystallographica Section B **30**, 1053 (1974).

[61] Wim J. A. Masskant, *On Helices Resulting from a Cooperative Jahn-Teller Effect in Hexagonal Perovskites*, in *Structure and Bonding*, Vol. 83 (1995), pp. 55–87.

[62] Andrew G. Christy, Ross J. Angel, Julian Hainest, and Simon M. Clark, *Crystal Structural Variation and Phase Transition in Caesium Trichlorocnprate at High Pressure*, Journal of Physics: Condensed Matter **6**, 3125 (1994).

[63] W. Castro Ferreira et al., *Pressure-Induced Structural and Optical Transitions in Luminescent Bulk Cs₄PbBr₆*, Journal of Physical Chemistry C **126**, 541 (2022).

[64] A. Boultif and D. Louër, *Powder Pattern Indexing with the Dichotomy Method*, J Appl Crystallogr **37**, 724 (2004).

[65] G. de La Flor, D. Orobengoa, E. Tasci, J. M. Perez-Mato, and M. I. Aroyo, *Comparison of Structures Applying the Tools Available at the Bilbao Crystallographic Server*, J Appl Crystallogr **49**, 653 (2016).

[66] A. L. Spek, *Single-Crystal Structure Validation with the Program PLATON*, J. Appl. Cryst **36**, 7 (2003).

[67] S. Ivantchev, E. Kroumova, M. I. Aroyo, J. M. Perez-Mato, J. M. Igartua, G. Madariaga, and H. Wondratschek, *SUPERGROUPS - a Computer Program for the Determination of the Supergroups of the Space Groups*, J. Appl. Cryst **35**, 511 (2002).

[68] F. Birch, *Finite Elastic Strain of Cubic Crystals\**, Physical Review **71**, 809 (1947).

[69] N. Sata, G. Shen, M. L. Rivers, and S. R. Sutton, *Pressure-Volume Equation of State of the High-Pressure (Formula Presented) Phase of NaCl*, Phys Rev B **65**, 1 (2002).

[70] A. L. Goodwin, D. A. Keen, and M. G. Tucker, *Large Negative Linear Compressibility of Ag₃[Co(CN)₆ ]*, PNAS **105**, 18708 (2008).

[71] T. Katsura and Y. Tange, *A Simple Derivation of the Birch–Murnaghan Equations of State (EOSs) and Comparison with EOSs Derived from Other Definitions of Finite Strain*, Minerals **9**, (2019).

[72] V. Svitlyk, D. Chernyshov, A. Bosak, E. Pomjakushina, A. Krzton-Maziopa, K. Conder, V. Pomjakushin, V. Dmitriev, G. Garbarino, and M. Mezouar, *Compressibility and Pressure-Induced Disorder in Superconducting Phase-Separated $Cs_{0.72}Fe_{1.57}Se_2$*, Phys Rev B **89**, 1 (2014).

[73] R. O. Agbaoye, P. O. Adebambo, and G. A. Adebayo, *First Principles Comparative Studies of Thermoelectric and Other Properties in the Cubic and Hexagonal Structure of $CsCdCl_3$ Halide Perovskites*, Computational Condensed Matter **21**, (2019).

[74] J. M. Chaves, O. Florêncio, P. S. Silva, P. W. B. Marques, and S. G. Schneider, *Anelastic Relaxation Associated to Phase Transformations and Interstitial Atoms in the Ti-35Nb-7Zr Alloy*, J Alloys Compd **616**, 420 (2014).

[75] Q. Tu, D. Kim, M. Shyikh, and M. G. Kanatzidis, *Mechanics-Coupled Stability of Metal-Halide Perovskites*, Matter **4**, 2765 (2021).

[76] S. Sun, Y. Fang, G. Kieslich, T. J. White, and A. K. Cheetham, *Mechanical Properties of Organic-Inorganic Halide Perovskites, $CH_3NH_3PbX_3$ (X = I, Br and Cl), by Nanoindentation*, J Mater Chem A Mater **3**, 18450 (2015).

[77] Y. Rakita, S. R. Cohen, N. K. Kedem, G. Hodes, and D. Cahen, *Mechanical Properties of $APbX_3$ (A = Cs or CH3NH3; X = I or Br) Perovskite Single Crystals*, MRS Commun **5**, 623 (2015).